# FPGA Acceleration of Image Reconstruction for Real-Time Photoacoustic Tomography

Zijian Gao[1], Yuting Shen[1], Daohuai Jiang[1], Fengyu Liu[1], Feng Gao[1,*] and Fei Gao[1,*]

*Abstract*—Photoacoustic (PA) imaging has been widely applied in both preclinical and clinical applications. With a significantly increasing number of data acquisition channels, fast and high-quality image reconstruction for real-time PA imaging is an open challenge in this community. In this paper, we propose a FPGA-accelerated method to achieve a much faster image reconstruction speed by 20~60 times compared with using CPU, with much-reduced system cost and power budget, from dozens of Watt (CPU) to 1~2 Watt (FPGA). Equivalently, the energy efficiency ratio (EER) is improved by ~1000 times. This FPGA acceleration method can be easily adapted to the most widely used algorithms, such as delay-and-sum (DAS) and its variants (e.g. DMAS, DAS-CF). We have performed *in-vivo* human finger experiments to demonstrate the feasibility and potential of the proposed method. To our best knowledge, this is the first study of accelerating PA image reconstruction based on FPGA platform.

## I. INTRODUCTION

Photoacoustic tomography (PAT) is a fast-developing imaging modality in recent years, which is based on the physics of photoacoustic effect. Biological tissue will generate ultrasound waves due to its thermal expansion followed by absorbing the energy of pulsed laser [1-3]. PAT combines the light absorption contrast with deep penetration of ultrasound for functional imaging (e.g. oxygen saturation) in deeper tissues. Thus, PAT can be applied in early screening of diseases, such as breast cancer and thyroid tumors [4, 5]. With the increasing requirement of clinical applications (e.g. 3D real-time imaging with affordable cost), the PAT system tends to be equipped with massively parallel multi-channel data acquisition (DAQ) modules, keeping real-time image reconstruction and low system cost simultaneously.

Some researchers proposed a delay-line-based time-multiplexing strategy, which can greatly reduce the number of DAQ channels to reduce system cost [6-8]. However, image reconstruction procedure is not accelerated in these work. Some other studies explored modified delay-and-sum algorithm, compressed sensing, and deep learning frameworks, mostly based on Graphics Processing Units (GPU) acceleration [9-14]. Some previous studies report PAT system based on Field Programmable Gate Array (FPGA) [15-17], which didn't focus on image reconstruction acceleration on FPGA platform.

Since its easy-to-implement structure, delay-and-sum, and its improved versions (e.g. DMAS, DAS-CF, DMAS-CF), are widely used in PA imaging [18-20]. In this paper, we propose to accelerate these classical delay-and-sum (DAS) algorithms for PA imaging based on FPGA platform. Compared with CPU-based platform, the proposed architecture in FPGA can increase the reconstruction speed by 20~60 times with much-reduced power consumption by more than 18 times. Equivalently, the energy efficiency ratio (EER) is improved by ~1000 times.

## II. METHOD

### A. Delay-and-sum algorithm

A typical PA imaging system employs an array ultrasound transducer with N channels. The relationship between the beamformed signal and each channel's PA signal can be derived as:

$$S_{DAS}(x,y) = \sum_{i=1}^{N} S(i, \tau(x,y,i)) \tag{1}$$

where $S_{DAS}(x,y)$ represents the pixel value of the position $(x,y)$ of the image plane. $S(i, \tau(x,y,i))$ is the PA signal's value of $i^{th}$ transducer's element at delay time $\tau(x,y,i)$, which is ultrasonic propagation time from the position $(x,y)$ to $i^{th}$ transducer's element. $\tau(x,y,i)$ can be calculated by

$$\tau(x,y,i) = \frac{l(x,y,i)}{c} \tag{2}$$

$l(x,y,i)$ is the distance between the position $(x,y)$ to $i^{th}$ transducer's element, and acoustic velocity is $c$, as shown in Fig. 1(a).

### B. System design of DAS architecture

According to Eq. (2), the $\tau(x,y,i)$ delay time depends on the distance $l(x,y,i)$ and ultrasonic velocity $c$. Then delay time $\tau(x,y,i)$ can be obtained, which is constant for a specific PA imaging system. Considering the sampling rate and points of DAQ card are also fixed, we can transform delay time into the serial number of the sampling point of the PA sensor data. Table data can be used to record each pixel's delay time by a serial number of sampling points, which has the same size as the finally reconstructed image. One example is shown in Fig. 1(b), if the transducer's element is in the upper left corner of the imaging region, we can see that the values of the table data at positions (1,1), (1,2), (1,3) are 2, 6, and 8. Then we can read values of sensor data from addresses 2, 6, and 8. These values a, b, c can be regarded as values of imaging plane at pixels (1,1), (1,2), (1,3). By repeating the same procedure above, the reconstructed PA image can be obtained.

Next, we design the hardware module to accelerate the speed of DAS algorithm. As shown in Fig. 1(c), since table data are fixed, we can allocate a space of Read-Only Memory (ROM) to store it. There are another two spaces of Random Accessed Memory (RAM), RAM1 and RAM2, to store PA sensor data and reconstructed PA image. The module controller can synchronize the data flow of the input of sensor data and the output of the imaging.

Firstly, the controller pushes sensor data into DAS Module, which can be stored in RAM1 with address 1~k (k depends on the length of DAQ's sampling point). Then we read the values of the table data in the ROM in order. The value of the table data is regarded as the address of RAM1, so that the corresponding value of sensor data $S(i, \tau(x,y,i))$ can be read and stored at the right position in RAM2. After going through all the table data in ROM, the output of imaging can be obtained. The workflow is shown in Fig. 1(d).

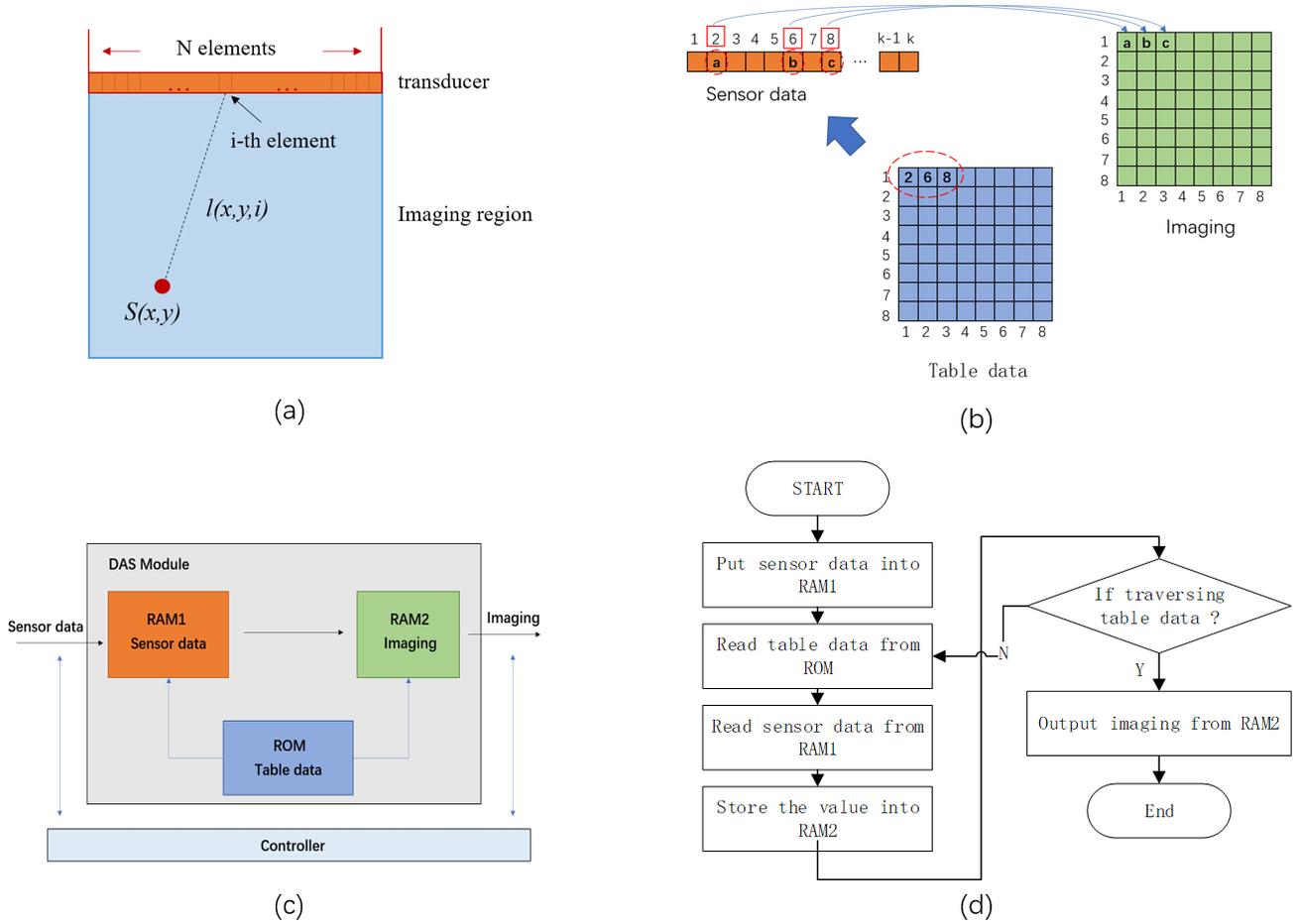

Fig. 1. (a) The concept illustration of the delay-and-sum algorithm. (b) Memory read and store for the delay-and-sum algorithm. (c) The design of the DAS Module. (d) The workflow of the DAS Module.

## C. Implementation of DAS architecture

As shown in Fig. 2(a), multi-channel sensor data can be processed simultaneously with multiple DAS modules to generate each-channel's image data. Then the final image is obtained by adding the image data of each channel. Mainstream FPGA devices have some independent IP cores. Therefore, we can use these IP cores to implement related variants of the DAS algorithm with little additional time consumption, based on the DAS hardware framework.

As shown in Fig. 2(b), the algorithm of DAS-CF can be achieved easily by slight modification. The DSP IP core can be used to perform data square operation, which costs one-clock cycle. Then, the Divider IP core can be applied to carry out divider operation with 20 clock cycles.

As for the algorithm of DMAS, we design a $\overline{\sqrt{x}}$ module [18], where $\overline{\sqrt{x}} = sign(x)\sqrt{|x|}$. In this module, the operation of fetching a sign can be done by determining the highest bit of the signed number. The absolute value operation can be done by changing the sign bit of the signed number. Then we can add the Cordic IP core into the $\overline{\sqrt{x}}$ module to perform the square root operation with 16 clock cycles. In addition, dividing by 2 can be done by moving the data one bit to the right. The hardware architecture of DMAS is shown in Fig. 2(c).

.

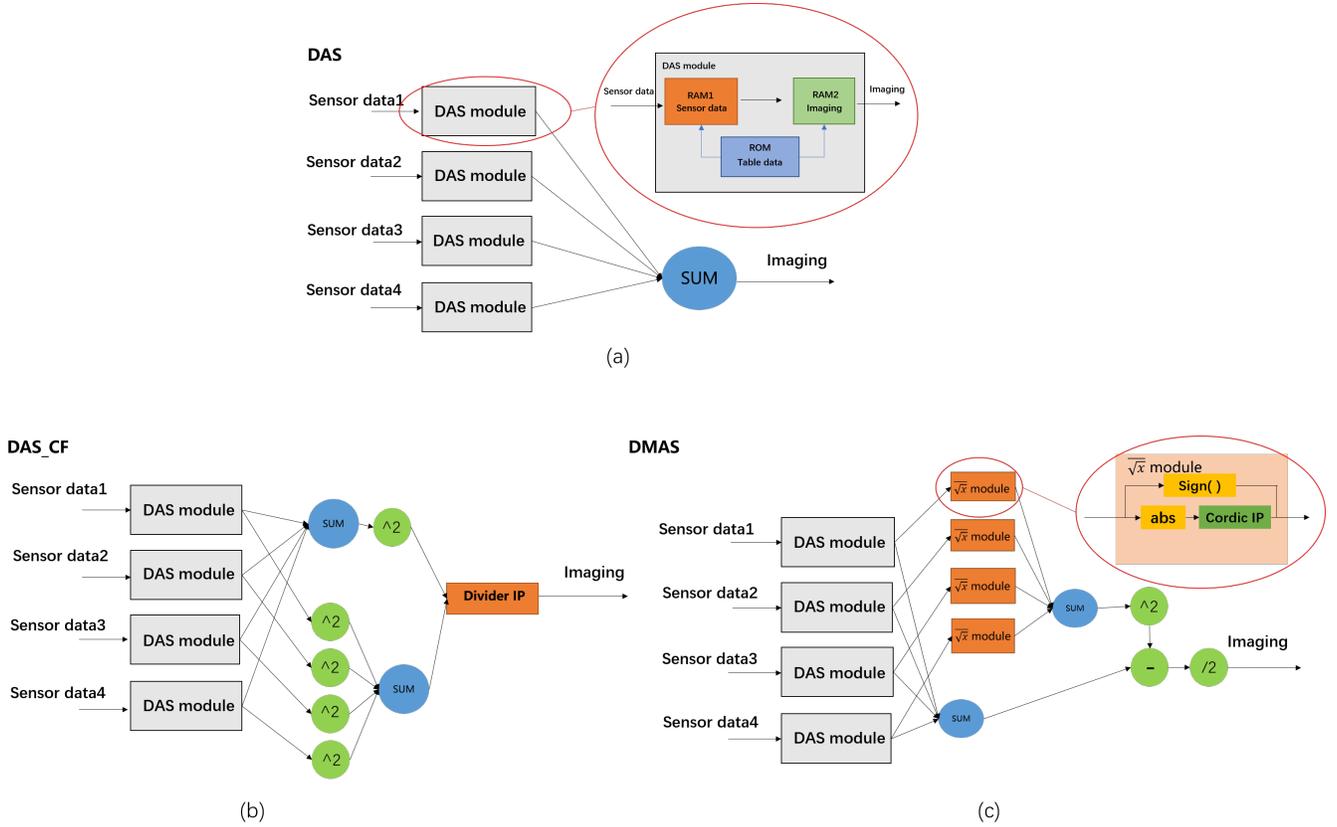

Fig. 2. The proposed hardware architecture of (a) DAS, (b) DAS-CF, and (c) DMAS algorithms.

III. EXPERIMENTAL RESULTS

To prove the feasibility of the proposed hardware-accelerated architecture, we performed the experiment at FPGA platform with *in-vivo* finger PA imaging data.

*A. Experiment setup*

The FPGA platform is Xilinx Zynq-7000, which has 200 MHz clock frequency, 53200 Look-up tables (LUTs) and 4.9 Mb block RAM. This platform receives PA sensor data from Analog-to-Digital Converter (ADC), and transmits PA images to PC, as shown in Fig. 3(a), where ADC and FPGA make up the DAQ card. There is a 128-channels ring-shape ultrasound transducer with a radius of 30 mm. The target lies in the center of the ring, which is illuminated by a pulsed laser to generate PA signals.

The region of interest (ROI) is in the center of the ring, which is a bounded square space of $20 \times 20$ mm$^2$. The frequency of the ADC's sampling rate is 40 MSPS, and the sound speed is 1500 m/s in a homogeneous medium. Considering limited FPGA's computing resources, the hardware architecture is designed in an 8-channel parallel mode.

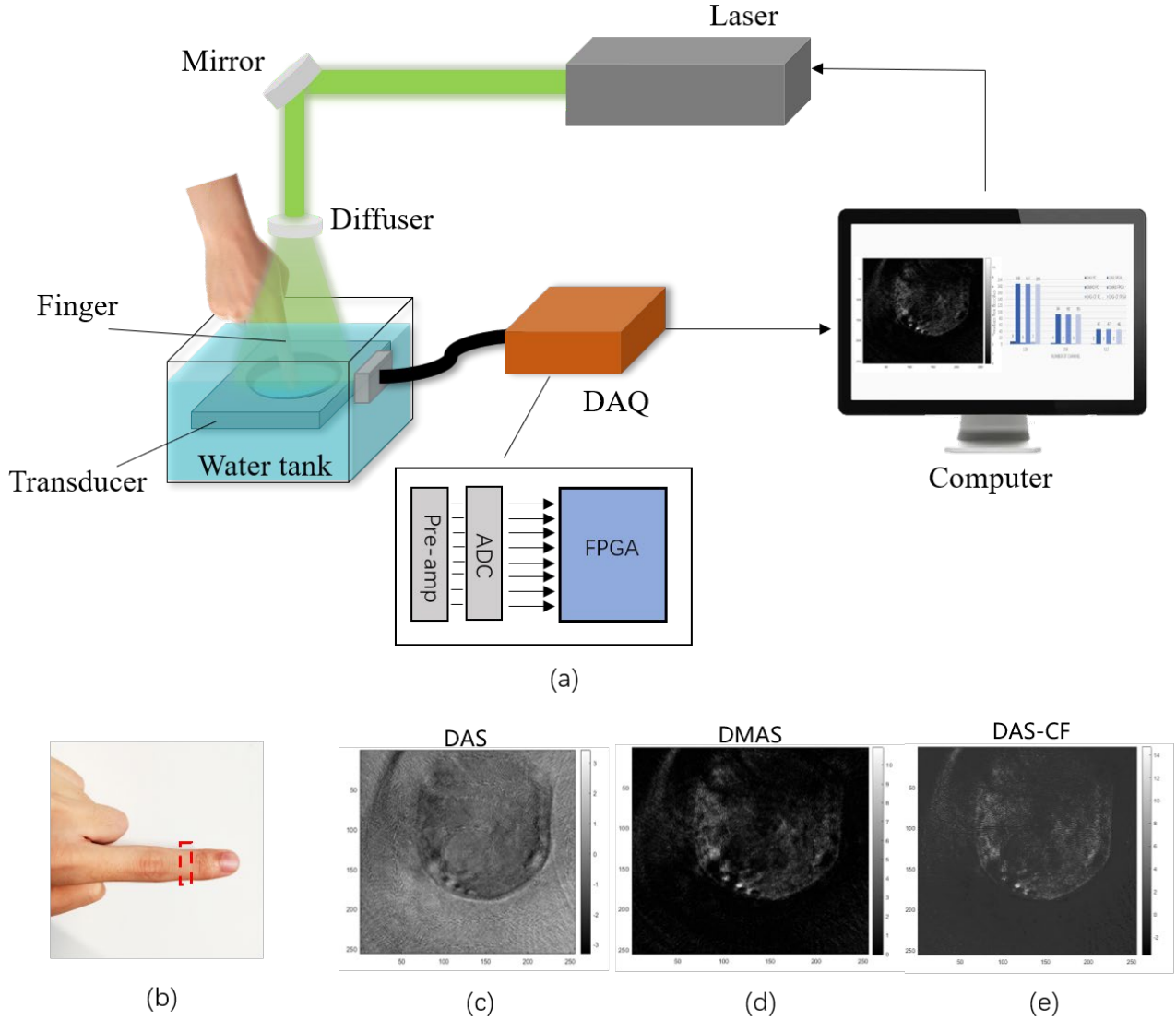

Fig. 3. (a) The experimental setup of photoacoustic imaging system with proposed FPGA acceleration module. (b) The image area of cross-section of human finger is indicated by the red dashed box. (c-e) The *in-vivo* cross-sectional PA imaging results of human finger with DAS, DMAS, DAS-CF.

## B. Results

We perform *in-vivo* finger experiments using our PA imaging system. As shown in Fig. 3(b), the red dashed box indicates the finger's cross-section. The proposed method can achieve image reconstruction of DAS, DMAS and DAS-CF algorithms with results shown in Fig. 3(c)-(e). In the data analysis, we calculate the time consumption of PA image reconstruction to compare the speed performance of FPGA and CPU. The CPU model is i7-1165G7, whose power consumption is about 28 W.

Firstly, we design experiment I with different ROI's sizes from $64 \times 64$ to $512 \times 512$ with fixed 128 transducer channels. The larger the image size, the more time it takes to read the table data. In TABLE I, with different imaging sizes, they show 30 to 60 times improvement in reconstruction speed. Then we plot these results in Fig. 4(a) in terms of frames per second (fps), achieving 47 fps for $512 \times 512$ image size using FPGA, compared with 1 fps using CPU.

Next, we design experiment II with different number of transducer channels from 128 to 512 with fixed ROI's sizes of $256 \times$

6256. Different number of channels can affect the entire imaging cycle, thus total time consumption. For example, our designed hardware architecture is based on 8 channels in parallel, which takes 32 imaging cycles to achieve 256 channels' PA image reconstruction. In TABLE II, the time cost with different transducer channels are illustrated, achieving speed improvement from 20 to 60 times, which is also shown in Fig. 4(b).

The Fig. 4(c) compares the power consumption of CPU and FPGA platform, which shows much less power consumption by more than 18 times (zynq-7020: ~1.5W, i7-1165G7: ~28W), equivalently achieving ~1000 times EER improvement.

TABLE I. TIME COST (s) OF EXPERIMENT I

| image size | DAS | | DMAS | | DAS-CF | |
|---|---|---|---|---|---|---|
| | CPU | FPGA | CPU | FPGA | CPU | FPGA |
| 64X64 | 0.013426 | 0.000418 | 0.017876 | 0.000433 | 0.01474 | 0.000423 |
| 128X128 | 0.033123 | 0.001394 | 0.052898 | 0.001431 | 0.042744 | 0.001498 |
| 256X256 | 0.114546 | 0.005327 | 0.192831 | 0.005328 | 0.137846 | 0.00533 |
| 512X512 | 1.344637 | 0.021 | 1.783402 | 0.021056 | 1.464202 | 0.021059 |

TABLE II. TIME COST (s) OF EXPERIMENT II

| channel | DAS | | DMAS | | DAS-CF | |
|---|---|---|---|---|---|---|
| | CPU | FPGA | CPU | FPGA | CPU | FPGA |
| 128 | 0.114546 | 0.005327 | 0.192831 | 0.005328 | 0.137846 | 0.00533 |
| 256 | 0.241636 | 0.010654 | 0.415223 | 0.010656 | 0.353224 | 0.01066 |
| 512 | 0.514915 | 0.021308 | 0.714216 | 0.021312 | 0.555601 | 0.02132 |

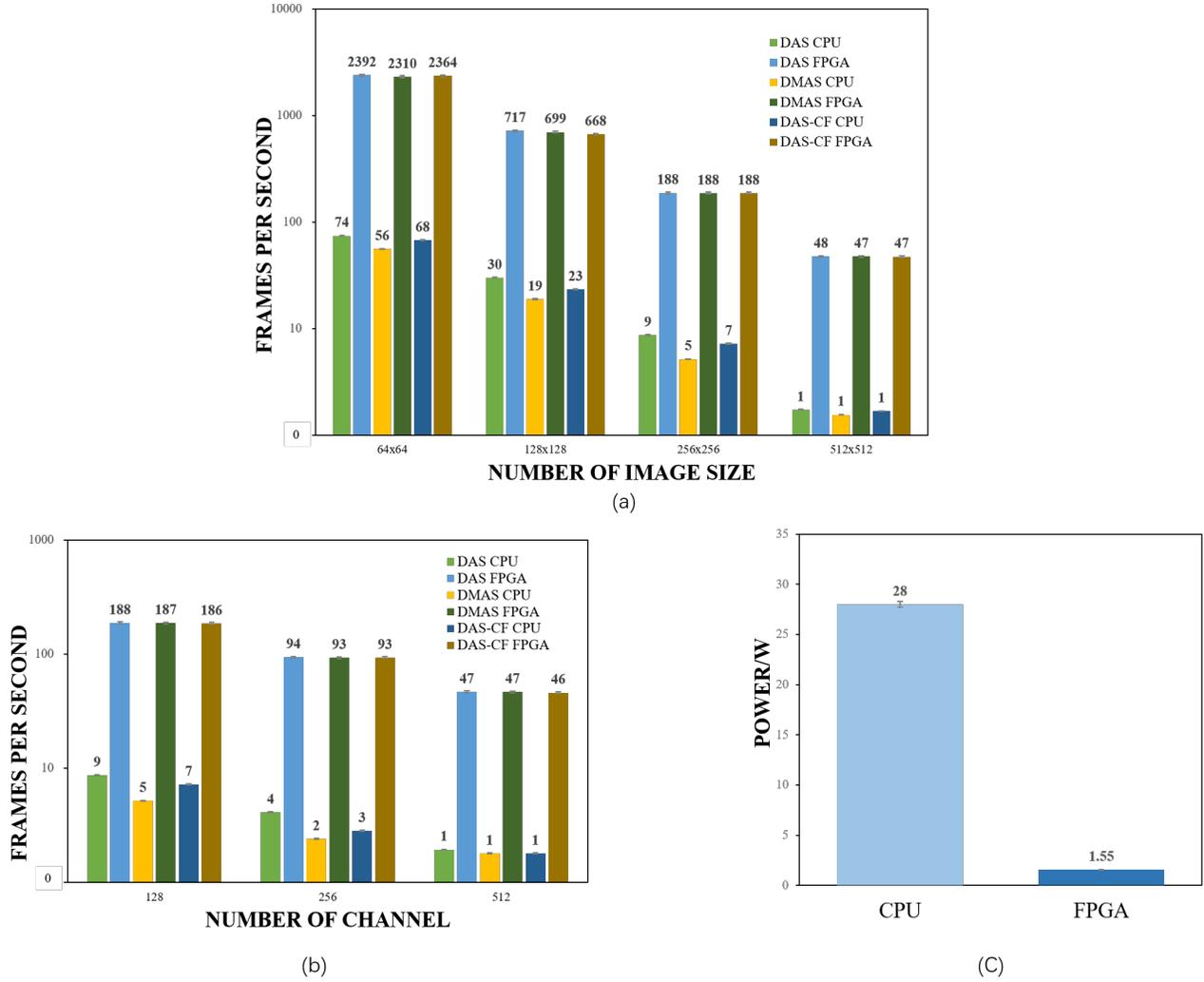

Fig. 4. (a) Frames per second of experiment I. (b) Frames per second of experiment II. (c) The power of CPU and FPGA.

## IV. Discussion and conclusion

In this paper, we propose a hardware-accelerated method for image reconstruction of PA imaging. We convert DAS algorithm to memory storage and access, so that we can use FPGA platform to accelerate the reconstruction speed. We also design hardware architecture of related DAS algorithm variants (e.g. DMAS, DAS-CF) by slight modification. *In-vivo* human finger imaging experiments are conducted. The result shows that our proposed FPGA acceleration method can achieve ~1000 times improvement in terms of EER compared with conventional CPU solutions. In future work, we will further optimize and integrate the FPGA platform into the PA imaging system for more clinical applications.

## V. Acknowledgement

This research was funded by National Natural Science Foundation of China (61805139), and United Imaging Intelligence (2019×0203-501-02).